# Learning from Physics Education Research: Lessons for Economics Education

Mark Maier* and Scott Simkins**

*Glendale Community College
Economics Department, Glendale, CA 91208, mmaier@glendale.edu
** North Carolina A&T State University
Academy for Teaching and Learning, 313 Dowdy Building, Greensboro, NC 27411
simkinss@ncat.edu

**Abstract.** We believe that economists have much to learn from educational research practices and related pedagogical innovations in other disciplines, in particular physics education. In this paper we identify three key features of physics education research that distinguish it from economics education research – (1) the intentional grounding of physics education research in learning science principles, (2) a shared conceptual research framework focused on *how* students learn physics concepts, and (3) a cumulative process of knowledge-building in the discipline – and describe their influence on new teaching pedagogies, instructional activities, and curricular design in physics education. In addition, we highlight four specific examples of successful pedagogical innovations drawn from physics education – context-rich problems, concept tests, just-in-time teaching, and interactive lecture demonstrations – and illustrate how these practices can be adapted for economic education.

**JEL Categories**: A2 – Economics Education and Teaching of Economics

**Note.** The authors gratefully acknowledge financial support from the National Science Foundation (DUE #04-11037). An earlier version of this paper was presented at the Southern Economic Association Annual Meeting in November 2006

## I. INTRODUCTION

In his comprehensive review of the evidence on teaching economics to undergraduates, William Becker (1997) noted that "economists are noticeably absent" from cross-disciplinary discussions focused on advancing teaching. Further, Becker suggested that while "much of the rest of higher education implements new approaches to teaching, traditional economists may be stuck in the rut of doing to undergraduates what their instructors did to them." (pp. 1353-54) More than a decade later, economists' involvement in cross-disciplinary dialogue and research on pedagogical inquiry and innovation is little changed, limiting economists' potential for developing effective teaching practices aimed at improving student learning.

However, we believe that there is much to learn by examining discipline-based education research and successful pedagogical innovations developed outside of economics and exploring their adaptability for economics education. By looking beyond traditional disciplinary boundaries we hope to gain fresh insights on how to improve both the teaching and learning of economics.

In this paper we focus specifically on what economic educators can learn from physics education research and the body of literature on student learning and classroom practice that has developed from this research over the last thirty years. Our choice of disciplinary focus is motivated by two factors. First, physics education research is arguably the most advanced of all science-related educational research, producing a deep and growing body of knowledge about student learning that has been used to develop innovative teaching methods, informative assessment techniques, and intentionally designed curricula aimed at improving student learning.[1] Second, physics shares similar learning challenges with economics, in particular the need for students to master key conceptual knowledge for modeling real-world behavior and develop facility with graphical, numerical, and mathematical representations. Physicists have spent decades investigating these issues, developing an extensive knowledge base of effective teaching practices that we believe has much to offer economists interested in improving student learning outcomes in economics courses.

In the first half of the paper we summarize some of the key features of physics education research, highlighting *differences in the approaches* that economics and physics education researchers take in their work and exploring how those differences influence new teaching pedagogies, instructional activities, and curricular design. Physics education research differs from economics education research in three important ways: (1) broad discipline-level support for research in physics education, (2) the "research-based teaching" framework that guides much of physics education research, and (3) the learning-theory foundation upon which current physics education research is based. We believe that these differences yield important insights that can be used to advance both economic education research and the classroom teaching of economics.

In the second half of the paper we explore how the physics education research framework can be adapted to economics, providing four specific examples of successful pedagogical innovations drawn from physics education – context-rich problems, concept tests, just-in-time teaching, and interactive lecture demonstrations – and illustrating how these practices can be adapted for economic education. Again, we believe that the intentional linkage of these practices with research on student learning provides a useful guide for improving student learning in economics.

## II. KEY FEATURES OF PHYSICS EDUCATION RESEARCH

### A. A Community Of Scholars Supporting Physics Education Research

Physics education research benefits from a large and well-established research community supported by a worldwide network of university-based physics education research groups (PERGs).[2] Most physics education programs include extensive graduate training in physics education aimed at college-level teaching and learning. While still small in number, graduates from these programs have a disproportionate effect on pedagogical innovation in physics education at the college level. In a number of cases, entire physics courses and curricula have been developed around the physics education research findings of a group of collaborative investigators.[3] Even though economics enjoys a long history of research on economic education [see Becker (1997)], much of this research has been limited to individual attempts to improve student learning, with little connection to a comprehensive economic education research agenda.

---

[1] See Stokstad (2001) and Wieman (2007) for overviews of the impact of physics education research. See McDermott and Redish (1999) for an extensive list of references (224), grouped into topical sections that is "meant to contribute to the establishment of a research base that can serve as a resource for ongoing improvement and enrichment of student learning in physics." (p. 755) The closest examples in economics are the *Journal of Economics Education* website [http://www.indiana.edu/~econed/] and the Research in Economic Education Database (REED).

[2] Heron and Melzter (2005) list more than 50 such groups, many of which are linked to Ph.D. programs in physics education. Some of the most well-known physics education research groups are at the University of Washington, the University of Maryland, the University of Colorado-Boulder, North Carolina State University, and the University of Minnesota.

[3] See, for example, Lillian C. McDermott, et. al. (1996) *Physics by Inquiry*, E. F. Redish (2003) *Teaching Physics with the Physics Suite*, Priscilla Laws (1997) *Workshop Physics*, and Sokoloff, Thornton, and Laws (1999), *RealTime Physics*.

Physics also enjoys a wide variety of outlets for presenting and publishing physics education research. In addition to multiple journals devoted to physics education, the American Association of Physics Teachers (AAPT), the discipline's professional society devoted specifically to physics teaching and physics education research, sponsors both a winter and summer annual meeting that includes hundreds of invited and contributed papers, as well as poster presentations.[4]

Finally, top scientists in the field, including Nobel-prize-winning physicist Carl Wieman, whose support for classroom-based instructional innovations in science education has attracted increasing attention, provide important leadership for physics education research.[5] The high stature of Wieman and other internationally renowned physicists carrying out physics education research provides high visibility and additional credibility for this work.

## B. Research-Based Teaching: Focusing On Student Learning Gaps

One cannot read far in the physics education research literature without coming across the term "research-based teaching," a thirty-year effort to build a common core of knowledge about student learning in physics. This cumulative body of knowledge intentionally informs current teaching practices and serves as a shared foundation for ongoing educational research within the discipline.[6] Central to physics education research is a consistent focus on understanding how students learn (or fail to learn) physics, in particular the conceptual roadblocks that hinder the development of deep, long-lasting learning.

In addition to the pre/post testing framework common in economic education research, physics education researchers often look closely at the actual process of learning in individual students, employing "think aloud" protocols where researchers observe and record (orally and visually) students engaged in solving a physics problem on a particular concept.[7] Historically, this effort began with dissatisfaction about student learning in introductory physics courses and the recognition that there were serious gaps between what instructors were teaching and what students were learning. Understanding the gap between what is taught and what is learned is a key focus of physics education research. The knowledge gained from this micro-level research often leads to innovative new pedagogies, teaching resources, and assessment processes that systematically improve student learning. If successful, these new practices are then re-tested at different institutions, and with different instructors, including those not directly involved in physics education research.[8]

To gain a sense of the way that physics education researchers approach their work and how it differs from the way economics education research is carried out we highlight a recent study by Lillian McDermott's University of Washington Physics Education Research Group [see Kautz, et. al., (2005a and 2005b)]. This study is representative of a broad body of physics education research focusing on close observation of student learning.

Beginning with a concept recognized as fundamental for understanding and using physics – the first law of thermodynamics – University of Washington PERG researchers interviewed 45 undergraduate students enrolled in an upper-division physics course to determine the prior knowledge that they were bringing to the class. All of the students had successfully completed a sophomore level physics course (and in many cases also a year of chemistry). These were students most professors would expect to be able to answer standard physics problems, especially in a multiple-choice format. Nonetheless, interviews revealed that most students could not apply a foundational physics concept to solve a basic applied physics problem.

---

[4] For highlights of past years' AAPT annual meetings, see http://www.aapt.org/Events/mtghighlights.cfm.
[5] See Wieman (2007).
[6] Redish (1999) describes the characteristics necessary for building a cumulative research base in physics education.
[7] See, for example, Podolefsky and Finkelstein (2007a).
[8] See, for example, Pollock and Finkelstein (2008).

Based on this initial qualitative inquiry, three new problems targeting this learning gap were written and administered to more than 1000 students in physics courses at four universities. Again, many of the students answered these problems incorrectly, a result that did not depend strongly on the type of course (algebra or calculus based) or timing of the assessment (before or after instruction). Responses were categorized in order to pinpoint the sources of student error. In turn, this information was used to design new instructional methods that specifically focused on closing the learning gaps identified in the study. Follow-up testing showed that the new pedagogy improved student learning of the targeted concept.

This example follows a physics education research protocol made popular by Lillian McDermott and the University of Washington PERG (which she led); it has been used in dozens of projects dating back nearly thirty years:

(1) administering problem-based physics tests on a core concept and carefully observing students solving these problems to determine student learning gaps;
(2) creating new instructional resources and techniques based on insights gained about these student learning gaps; and
(3) testing the effectiveness of the new teaching methods and materials in closing the gaps

Physics education researchers' consistent, long-term focus on the *process* of observing student learning, identifying learning gaps, and developing instructional strategies to address them has created a growing cumulative knowledge base that is widely shared among physics education researchers. In turn, this common knowledge base has served as the foundation for broad-based instructional and curricular reforms in physics education.

Complementary efforts in economics are difficult to find. The most similar work in economics education was conducted by Strober (1992, 1997), who analyzed videotape of economics students as they solved problems. However, Strober's work remains largely unknown and has had little impact on the teaching of economics across the discipline. More recently, work on "threshold concepts" by Davies and Mangan (2007) has focused attention on key economic concepts that serve as both roadblocks and gateways to learning in economics, but the impact of this work on classroom teaching remains to be seen.

# C. Grounding Teaching Practices In Learning Sciences Research

Physics education researchers share a common theoretical framework characterizing important aspects of student learning, a framework that relies heavily on a small set of robust research findings from the learning sciences literature that are associated with deep and long-lasting learning. Three key learning issues derived from this literature [see Bransford, Brown, and Cocking (1999)] are central to most physics education research: (1) identifying and directly addressing student preconceptions; (2) developing students' metacognitive skills; and (3) increasing students' ability to transfer knowledge to new situations.

### Identifying and Addressing Student Preconceptions

The starting point for nearly all physics education research is identifying how students misperceive core concepts, both before *and often after* successfully completing a traditional physics course.[9] One of the key reasons for this intellectual disconnect is that students bring their own mental models to our classrooms, mental models that retard the development of new knowledge and are resistant to change. As a result, students often maintain naïve views of scientific concepts even after completing multiple science courses.[10]

---

[9] Many times, these misperceptions continue even after *advanced* study in the field. See McDermott and Redish (1999) for a wide variety of resources on this topic.
[10] See Harvard-Smithsonian Center for Astrophysics, (1997) *A Private Universe*, for an example of the persistence of

The notion that instructors need to identify and address students' current understanding is central to the types of resources and pedagogical innovations that have been developed in physics education. Most notably, physicists have been at the forefront in developing "concept inventories," short multiple-choice tests designed to reveal persistent student preconceptions. The most famous of these is the Force Concept Inventory (FCI), developed by Halloun and Hestenes [see Halloun and Hestenes (1985a, 1985b) and Hestenes, Wells, and Swackhamer (1992)]. Concept inventories such as the FCI use semi-realistic situations and everyday speech embedded in distracter answers derived from common student preconceptions identified by previous physics education research. The FCI has proven to be a valuable instrument for uncovering persistent misconceptions.[11]

Many physicists are stunned to discover that students who have completed their physics courses are unsuccessful at answering questions on the FCI. As Harvard physicist Eric Mazur reports, "…the results of the test came as a shock… Clearly many students in my class were concentrating on learning "recipes" or "problem-solving strategies" as they are called in textbooks, without considering the underlying concepts." [Mazur (1997), pp. 4-7] This result led Mazur to revise his instructional approach to intentionally address student preconceptions using an innovative set of teaching strategies, *Concept Tests* and *Peer Instruction*, that has produced measurable gains in student learning [see Mazur (1997), Crouch and Mazur (2001), and Crouch, Watkins, Fagen, and Mazur (2007)].

In economics education, the importance of students' prior experience on student learning has been previously recognized by Saunders [in Saunders and Walstad (1998)]. However, unlike physics education research, there has been little or no research in economics attempting to systematically identify and catalog student preconceptions and develop instructional methods to intentionally address them. Perhaps this is because there is no consensus on what constitutes an appropriate list of core concepts for introductory economics courses [Hansen, Salemi, and Siegfried (2002)]. Nonetheless, concept inventories hold promise as a way to identify student preconceptions that persist even after studying economics – even if economists may not select the same core concepts as most important. Once these preconceptions are identified, pedagogical strategies and curricular materials can be developed to confront students with these preconceptions directly and change their underlying mental models.

## Promoting Students' Metacognitive Skills

Another important area for improving student learning is helping students understand and monitor their own thinking processes, or to become more metacognitive. As Redish (2003) points out, students are generally unfamiliar with this type of thought process.

> The key element in the mental model I want my students to use in learning physics appears to me to be *reflection*—thinking about their own thinking. This includes a variety of activities, including evaluating their ideas, checking them against experience, thinking about consistency, deciding what's fundamental that they need to keep and what is peripheral and easily reconstructed, considering what other ideas might be possible and so on. My experience with students in introductory classes – even advanced students – is that they rarely expect to think about their knowledge in these ways. (p. 62)

For deep and durable learning to occur, students need to recognize the gaps in their understanding and must be provided with multiple opportunities to both identify and address those gaps. In addition, students must be trained to intentionally reflect on their own thought and problem-solving processes by asking questions about *how*

---

naïve models and the difficulty in changing those views via classroom teaching.
[11] Concept inventories are now available in most STEM fields. See *Concept Inventory Central* (2008) for a comprehensive listing. See also Richardson (2004) for information on using concept inventories for uncovering student misconceptions.

problems were solved. In particular, students need to continually ask themselves: W*hat information do I know about this problem/issue/question?, What information do I need to find?,* and *How would I obtain that information?* Without the development of metacognitive skills – explicit recognition by students of their own thought processes, including contradictions between their observations and their implicit mental models – student thinking is likely to be surface-level and episodic.[12]

## Developing Students' Ability To Transfer Knowledge To New Situations

The third area for improving student learning is helping students develop the skills necessary to transfer knowledge beyond the context in which it was taught. As McDermott (1991) notes:

> What the instructor says or implies and what the student interprets or infers as having been said or implied are not the same. There are often significant differences between what the instructor thinks students have learned in a physics course and what students may have actually learned. (p. 303)

Many of the problems identified by physics education researchers deal with students' inability to transfer knowledge to new contexts. Mazur (1997), for example, notes that most physics students quickly become proficient at memorizing physics formulas, applying them in a "plug and chug" fashion. Their problem-solving strategies rely on novice-like algorithms rather than the structurally connected and organized knowledge maps of experts. As a result students are often unable to transfer basic principles and concepts to environments unlike those in which they were originally presented.

Physics education researchers have also re-examined what it means for students to apply understanding in new situations. For example, students may be able to transfer knowledge by *solving a problem* in a new context, but they may not carry over this knowledge to *learn* in a new context as they move through a traditional physics curriculum.[13] For example, students might learn how to correctly use Newton's third law in a variety of contexts but fail to transfer the concept into subsequent learning about multibody problems. Physics education researchers such as Joe Redish conclude that transfer for learning – in addition to problem-solving – is best accomplished when instruction explicitly asks students to connect their understanding to prior knowledge and to be metacognitive about their learning [Hammer et. al. (2005) p. 115]

In economics, the importance of student transfer of economics principles and concepts was explicitly recognized by Saunders [in Saunders and Walstad (1998)] and is implicit in techniques such as the case method and problem-based learning, which have been incorporated in economics instruction in recent years [see Velenchik (1995), Carlson and Velenchik (2006), Meyers (2008), and Higher Education Academy Economics Network (2008)]. However, beyond the underlying recognition that transfer matters for student learning, economic educators have failed to intentionally integrate key insights about knowledge transfer from the learning sciences when developing new classroom teaching practices.

Physics education researchers, conversely, have maintained a steady focus on the transfer issue and have systematically worked to develop teaching resources and practices that directly address this challenge. For example, a recent University of Colorado Physics Education Research Group study examined transfer of student knowledge about water waves and sound waves to an understanding, by analogy, of electromagnetic waves [Podolefsky and Finkelstein (2007b, 2007c)]. In economics, a similar study might look at ways in which

---

[12] Metacognition can be promoted in many ways, including specific classroom assessment techniques such as the one-minute paper, which has previously been investigated by economic educators (Chizmar and Ostrosky, 1998). However, metacognition is more often woven throughout other pedagogical techniques such as the McDermott tutorials noted earlier.

[13] A valuable discussion of issues surrounding transfer of knowledge is provided by Schwartz, Bransford and Sears (2005).

Maier and Simkins, *Learning from Physics Education Research*, June 2008    6

students transfer knowledge about price elasticity of demand from one good or service to another, but may or may not be able to transfer their understanding to other types of elasticity.

## III. LESSONS FOR ECONOMISTS

Physics education research is characterized by a systematic approach to uncovering, explaining, and addressing student learning gaps, focusing primarily on students' learning *processes* rather than specific knowledge acquisition. The result of this learning-centered approach to physics education is the development of a cumulative knowledge base of *how students learn physics* that drives both educational research and pedagogical innovation in physics. Results from the learning sciences provide a common theory-based foundation for physics education research that further supports cumulative knowledge building within the discipline.

The work of Lillian McDermott, Joe Redish, and others within the physics education community have helped to outline a valuable framework for building a common disciplinary knowledge base. The key features of this framework are outlined below.

(1) Begin with a demonstrated knowledge gap with respect to a key concept in the field
(2) Develop a deep understanding of student learning processes with respect to this concept
(3) Design learning resources and pedagogical methods to directly address the learning gap based on the observed student learning processes
(4) Assess student learning after integrating the new learning resources and pedagogical methods and compare to previous results
(5) Check for robustness of results across instructors, institutions, and students.
(6) Add to the cumulative physics education knowledge base

Economists will notice some similarities between the process outlined above and that followed by economic education researchers, yet qualitative differences between the two fields are apparent.

This point is most clearly illustrated in the way that research knowledge develops and grows within the discipline of physics education. Indeed, a defining characteristic of physics education research is the intentional, long-term process of cumulative knowledge building. To be sure, economists build on the work of others, but too often this work is difficult to place in a broader economics education research agenda in the same way that is possible with physics education research.

Educational research on economic experiments provides perhaps the best example of a sustained community of researchers working on a common pedagogical framework within economics education, but even here there has been neither an intentional connection to learning theory nor a systematic, discipline-wide building of evidence.[14] As is the case with much economic education research, the issues and topics often reflect the interests of individual researchers rather than the intentional focus of the discipline. As a result, it is difficult to identify an ongoing thirty-year research agenda within economics of the types developed by physics education research groups.

While economic education researchers have taken a lead role in applying sophisticated statistical tests to assess the impact of instructional innovations on student learning [Becker (2004)], physics education research offers a complementary educational research model for building a coherent core body of knowledge explicitly aimed at improving student learning. This model has produced innovative pedagogical strategies, instructional activities, and curricular reforms that have been shown to improve student learning in physics.

Based on current and previous research we have conducted as part of two National Science Foundation projects (NSF DUE-0411037 and DUE 00-88303) we believe that physics education research provides valuable insights for economic education researchers and classroom teachers. Specifically, we believe it is worth experimenting with adaptations of successful physics education

---
[14] For a survey of the use of experiments in economic education see Hazlett (2006, pp. 21-38).

pedagogical innovations in economics and testing their usefulness in promoting student learning.

Among the most promising physics-developed pedagogical innovations are: (A) context-rich problems, (B) concept tests/peer instruction, (C) just-in-time teaching (JiTT), and (D) interactive lecture demonstrations (ILDs). Below we briefly describe each of these innovations and illustrate how they can be adapted for use in economics.

## A. Context-Rich Problems

Research by the University of Minnesota PERG explicitly addresses the issue of knowledge transfer to new contexts and advocates a technique for writing questions that promote knowledge transfer called *context-rich problems* [Heller, Keith, and Anderson (1992); Heller and Hollabaugh (1992)]. In order to engage student interest, link to previous student experiences, and provide guidance about the type of writing needed to answer the question, context-rich problems begin with "You…" and then place the student in a specific situation. For example, *You have been asked by your* [roommate, boss, relative, etc.] *to complete a task* [helping write a novel, explain why something happens in the home, consult with a moviemaking company, etc.].

The most important contribution of the context-rich approach is the attention to what is included or excluded in the problem. In designing a problem, instructors consider whether the problem should:

(1) include a diagram (or not);
(2) include excess data (so that the student needs to select the relevant information);
(3) exclude information that should be common knowledge or could be calculated based on common knowledge; and
(4) specify a target variable (or not).

Most context-rich problems are made more complex because of these decisions, yet they reflect the challenges common in real world situations. Through careful and intentional scaffolding students can learn to transfer learning to new situations by solving increasingly less structured questions in which diagrams, data, or target variables may or may not be present.[15] Consider the physics-based context-rich-problem below, taken from the University of Minnesota physics education research and development web site [http://groups.physics.umn.edu/physed/Research/CRP/crintro.html]:

---

**Context-Rich Problem Example – Physics**

*Because of your physics background, you landed a summer job as an assistant technician for a telephone company in California. During a recent earthquake, a 1.0-mile long underground telephone line is crushed at some point. This telephone line is made up of two parallel copper wires of the same diameter and same length, which are normally not connected. At the place where the line is crushed, the two wires make contact. Your boss wants you to find this place so that the wire can be dug up and fixed. You disconnect the line from the telephone system by disconnecting both wires of the line at both ends. You then go to one end of the line and connect one terminal of a 6.0-V battery to one wire, and the other terminal of the battery to one terminal of an ammeter (which has essentially zero resistance). When the other terminal of the ammeter is connected to the other wire, the ammeter shows that the current through the wire is 1 A. You then disconnect everything and travel to the other end of the telephone line, where you repeat the process and find a current of 1/3 A.*

---

Unlike non-contextual plug-and-chug physics problems that require only the selection of the proper formula and the insertion of numerical values for the variables, this physics problem requires students to critically think through a series of questions that parallel the context-rich problem structure: What is being asked? What information do I have? What information do I need? How will this information help me solve this real-world problem?

---

[15] A related issue identified by the physics education research group at the University of Colorado-Boulder is the difficulty that students encounter when moving from one representational form to another. For example, students often have difficulty moving from a graph to an equation or from a quantitative example to a qualitative explanation for the same problem. The UC-Boulder PERG is actively focused on helping students understand the benefits of multiple representations and select representations that will be most beneficial in specific contexts [Kohl and Finkelstein (2005)]. This provides another potential model that could be adapted to economics instruction and could form the basis for additional economic education research.

Now consider how the following traditional economics problem might be translated into a context rich problem.

*Calculate the present value of $10,000 received in ten years. Assume a discount rate of 4%.*

This type of economics question resembles the typical formula-based problems students regularly face in traditional physics classes. It includes all the information needed to solve the problem except for the formula, so students need only plug in the right numbers without making any discriminatory judgment. The problem is, students rarely will encounter this type of problem in the real world. As a context-rich alternative, the problem can be rewritten as:

**Context-Rich Problem Example – Economics**

*You and your brother have inherited a US government bond that will pay $10,000 in ten years but will not pay any interest before then. You agree to share the bond equally, but your brother would like to receive his one-half of the bond's value now. How will you explain to your brother, who has not studied economics, how much you should pay him for his share?*

In this case, the problem has an attention-grabbing, yet plausible context, and provides the student with an audience to whom the answer should be addressed. It is complex enough to be worked on by a group of students, there is missing data (the discount rate), and the specific target variable (the present value) is missing.

Of course, the context-rich problem approach is not the only teaching strategy to incorporate real-world or context-rich situations as the basis for problem solving. Problem-based learning and case study methods are two examples drawn from the economics education literature that display similar characteristics.[16] However, the context-rich problem approach offers a *formal structure* for transforming traditional problems in ways that facilitate transfer, an intentionality and ease of use that does not exist with other "real world" approaches.

The problem above illustrates that basic context-rich problem structure. Additional context-rich problems, perhaps used as follow-up work, could further encourage students to transfer recently gained knowledge to new situations. For example, students could be asked to solve bond price problems with different parameters, solve present value problems for other financial instruments, or apply the concept of present value in unexpected situations such as the value of environmental regulations.

## B. Concept Tests / Peer Instruction

One of the most widely adopted physics education pedagogical strategies is the peer instruction/concept test model developed by Harvard physicist Eric Mazur. This model is a variation on the cooperative learning technique called think-pair-share in which traditional lectures are stopped after five to seven minute segments to present a challenging multiple choice question, with distracters chosen to replicate common student errors identified by physics education research.[17] Students first consider the question and answer it individually, either with colored cards or personal response systems. The results of the polling are shown to the class, after which students consult with a classmate (peer instruction) and then are repolled.[18]

---

[16] See Velenchik (1995), Carlson and Velenchik (2006), and the resources at
http://www.economicsnetwork.ac.uk/handbook/pbl/.
[17] The think-pair-share technique is summarized in Millis and Cottell (1998) and Barkley, Cross and Major (2005).
[18] See http://galileo.harvard.edu/galileo/lgm/pi/ and University of Colorado adaptation at:
http://www.colorado.edu/physics/EducationIssues/ct_local.html Chemistry applications are discussed at
http://www.flaguide.org/cat/contests/contests1.php and

Evidence from Harvard and other institutions suggests that, in comparison with traditional lectures, peer instruction coupled with concept tests improves student conceptual understanding and problem-solving skills. Mazur has published a set of guidelines for concept tests, as well as a library of hundreds of ready-to-use short, multiple-choice questions, each dealing with a single fundamental physics concept and requiring a simple qualitative answer.[19]

An example of one of Mazur's concept test questions, focusing on a specific force concept, is illustrated below.

---

**Concept Test Example - Physics**
**Eric Mazur, Harvard University**

*The Levi Strauss trademark, shown on the left in the figure below, shows two horses trying to pull apart a pair of pants. Suppose Levi had only one horse and attached the other side of the pants to a fencepost, as illustrated on the right in the figure below.*

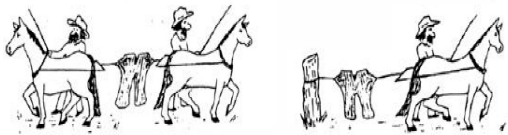

*Using only one horse would:*

*(a) cut the tension on the pants by one-half.*
*(b) not change the tension on the pants at all.*
*(c) double the tension on the pants.*

---

As this example makes clear, concept test questions are deceptively simple, yet provide immediate formative assessment of learning to both students and instructors. This knowledge can be used immediately to probe further for student misconceptions, provide additional examples, or

---

chemistry ConcepTest examples are available at: http://www.chem.wisc.edu/~concept/
[19] http://galileo.harvard.edu/

initiate small group exercises in class. Used together, concept test questions and peer instruction provide a "teachable moment" that differs from simply asking questions and promotes student-student interaction, even in large lecture classes.

Recent work in economics education on the use of personal response systems [Salemi (2008) and Elliot (2003)] and cooperative learning [Bartlett (2006)] shows that economists are beginning to experiment with formats similar to concept tests. However, economic educators have focused on practical issues such as solving student free-rider problems rather than understanding students' learning processes, prior mental models, and ability to transfer knowledge to new situations.

An example of a potential economics concept test question that examines not only conceptual issues but also issues related to student learning appears below.

---

**Concept Test Example – Economics**

*You read in the newspaper that U.S. imports from Japan have fallen due to a recession in the U.S. As a result, you predict that the value of the U.S. dollar, relative to the Japanese yen, will:*

   *(a) rise because fewer dollars are supplied to the foreign exchange markets.*
   *(b) fall because the demand for goods and services from Japan is reduced.*
   *(c) not change because U.S. consumers will buy more U.S. goods to offset the reduction in imports from Japan.*

---

This concept test question was chosen based on the anecdotal observation that students often confuse the currency market with goods and services markets, erroneously choosing answer (b) or (c).

When used in class, concept test questions provide instant feedback on student learning and allow for real-time instruction to correct lingering misperceptions or misunderstandings. Economics educators would benefit from a library of concept test questions such as these, focusing on the most common conceptual misunderstandings that students have in economics. Such a project is ideal for National Science Foundation (NSF) funding, further extending a pedagogical technique that has been shown to be effective in a wide variety of disciplines, often with NSF support.

## C. Just-in-Time Teaching

Just-in-Time Teaching (JiTT) is an instructional innovation first developed by physics educators [Novak, Gavrin, Christian, and Patterson, (1999)] and adapted for use in economics by Simkins and Maier (2004). JiTT exercises focus on material that will be covered in the next class and are aimed at getting students to come to class prepared, as well as providing information about the level of students' understanding of a particular concept or topic prior to class. Between classes, students complete brief, carefully constructed exercises – covering material that has not yet been covered – and submit them electronically using course management software, a web-based form, or simple email by a pre-assigned time a few hours before class.

Once submitted, instructors review students' JiTT responses and use them to organize and structure the upcoming classroom session – hence the "just-in-time" label. Excerpts from students' submissions are presented at the start of class as the basis for discussion, replacing the traditional lecture, and are used to develop follow-up exercises that groups of students work on in class. In addition, JiTT exercises often include a reflective, *"What is still unclear after reading the material for this JiTT exercise?"* question, encouraging students to develop metacognitive reading practices and providing instructors with yet another opportunity to get a clearer understanding of student questions and potential misunderstandings prior to class.

JiTT exercises also provide an incentive for students to complete reading assignments prior to class, helping to offset a persistent complaint of faculty members and leading to more student engagement in classroom activities and discussion.

A sample JiTT question from physics is listed below. This specific question asks students to employ concepts verbally, although JiTT questions can include numerical calculations as well.

---

**Sample JiTT Assignment – Physics**

*A person with good vision finds that she cannot focus on anything underwater. However, plastic goggles with "lenses" that are <u>flat</u> plastic disks allow her to see the fish clearly. Please explain how this can be.*

---

Note that this JiTT exercise focuses on a specific physics concept, yet presents that concept within a real-world example that encourages students to connect new theoretical knowledge to their current mental models. We have found that the most successful JiTTs share this characteristic.

Consider the following example of a Just-in-Time Teaching question, used early in a Principles of Economics course, focusing on student confusion about the relative importance of fixed and marginal costs in economic decision-making.

---

**Sample JiTT Assignment – Economics**

*Last year my wife and I made plans to take our family to the beach for the Labor Day weekend, accompanied by another family. Each family paid a non-refundable beach rental payment of $350 a couple of months prior to the trip. As Labor Day approached we watched the weekend weather report with growing interest. The weather forecaster was predicting rain for the entire weekend! As we packed up the car to go to the beach, I asked my wife if perhaps we should stay home for the weekend, rather than going to the beach. After all, we had recently moved and needed to unpack (and paint). She responded, "We've already paid $350 for the beach rental, of course we're going to the beach!"*

---

*Was my wife's argument "rational," in an economic sense? Why or why not?*

---

Below is a sample of student responses to this JiTT question, which were selected to illustrate the range of understanding in the overall set of responses. The four responses were displayed at the start of class along with the indicated discussion question, providing the basis for a collaborative learning exercise in class. Note that the exercises can also provide opportunities for formative assessment of student writing, both in terms of mechanics and organization.

---

**Student Responses (spelling unchanged)**

*Which of these responses comes closest to the argument that an economist would make? After selecting one of the responses below, how might you improve the answer, based on your understanding of the material you read for today's class?*

**Student #1:** *I think I would decide to go to the beach, rather than staying home to unpack and paint. I think the wife's decsion to go to the beach was a good economic decision. To me, the additonal benefit of the activity is greater than the additonal cost of the activity. I feel like if I have already paid for a $350 trip, if I fail to go I will be wasting money. There will always be more time to paint and unpack. Even if it is raining at the beach, the family might have the additional benefit of spending some quality time together. My opinion of wasting money is alot more important to me than wasting time. I would rather waste time over money any day.*

**Student #2:** *No because you will spend most if not all of your vacation inside due to the rain. That time could be spent working on your house. You also save money because you're at home versus being on vacation where you'd spend money on keepsakes, gifts and other needless items. By not going, you also save money on gas and food. The marginal cost of going to the beach outweigh the marginal benefits and thus it is not rational to go the beach.*

**Student #3:** *I believe that your wife's decision was a rational one in an economic standpoint because the non-refundable $350 deposit could have been used for some other type of activity for the new home. Even though the weather did not suit your standards it would be irrational to waste $350 by not taking the vacation.*

**Student #4:** *Well i personally feel that her argument is very rational, because the trip was already paid for and the trip was going to be restfull something that was more worth her time.*

---

The major innovation with Just-in-Time Teaching, relative to other teaching tools used to promote student preparation for class (e.g. quizzes at the start of class), is the potential attention to metacognition and transfer in follow-up activities. As in the example above, excerpts from students' submissions can be presented during the class as the basis for discussion, replacing the traditional lecture, or can be used to develop collaborative learning exercises that groups of students work on in class while the concepts are still fresh in students' minds.[20]

Students' JiTT responses also provide faculty members with valuable information about the concepts that are most difficult for students to understand, making visible students' thinking processes and preconceptions, as illustrated in the responses above. This information allows faculty members to target "just-in-time" classroom activities where they will provide the greatest benefits for student learning at a time when they can most profitably influence that learning. Instructors often employ "reading quizzes" that attempt to achieve some of the same goals as JiTT exercises. Yet, the information about student learning provided by these quizzes often comes

---

[20] See also Henderson and Rosenthal (2006) for a discussion of employing JiTT-like "reading questions" to promote student reading prior to class.

too late, when students and faculty members have moved on to the next course topic.[21]

## D. Interactive Lecture Demonstrations

Interactive Lecture Demonstrations (ILDs) illustrate the intentional integration of learning theory into classroom pedagogy. Developed by Sokoloff and Thornton (1997) and extended by Edward F. (Joe) Redish and his physics education colleagues at the University of Maryland PERG, ILDs require students to relate the concepts under investigation to their existing personal mental models and carry out demonstrations or experiments that are likely to challenge those models.

At first glance, ILDs look like simple classroom worksheets. However, they are carefully constructed to focus on a single conceptual topic, address students' mental models, and build metacognitive skills in an intentional, systematic process. The demonstration may take many forms, including traditional lab work or a thought experiment. The key is that students must record their prior experience with similar activities, explicitly describe models that might explain the demonstration, and then reflectively articulate how they have resolved differences between their initial understanding and the results of the demonstration.

The physics example below, taken from the University of Maryland PERG online resources, illustrates the multi-stage learning process characteristic of ILDs.[22]

---

[21] Unlike paper-based quizzes, personal response systems (or "clickers") allow for immediate quiz feedback but lack the rich insights into student thinking provided by JiTT exercises.
[22] See http://www.physics.umd.edu/perg/ for a comprehensive collection of physics education resources, including ILDs.

---

**ILD Example – Physics**

*Making a Model: Thinking about Electric Force*
*© University of Maryland Physics Education Research Group*

*1. Personal Experiences*

*What experiences have you had that you attribute to static electricity? Considering these experiences, can we generate a description of what's happening that is consistent with the Newtonian Synthesis we built in the first term?*

*2. Modeling Skills Practice*

*Do your general rules account for the observations of the paper bits sticking to the glass and the balloon sticking to the wall? You may feel sure that there are two types of charges that can attract or repel, but is that the only model your observations support? Can you rule out another model?*

*3. Experiment*

*Get a piece of tape (2 or 3 inches) and fold over a little bit of one end. Stick the tape to your desk with the folded end sticking out over the edge of the desk. Write the letter "B" on the tape. Now get another piece of tape and fold the end in the same way. Put the second tape directly on top of the first and write the letter "T" on it. Pull both tapes off the desk, and then holding the folded ends in opposite hands, pull the two tapes apart. With your neighbor, observe and record what happens when you:*

  *(a) put two "T" tapes near each other.*
  *(b) put two "B" tapes near each other.*
  *(c) put a "B" and a "T" tape near each other.*

*Does distance matter? How? Use your observations to help you evaluate the two different models and to add to or modify them as necessary.*

*4. Resolution*

*How does the model you've chosen explain the attraction of paper bits to a comb and the rubbed balloon sticking to the wall?*

*5. Reflection (Metacognition)*

*Instead of simply giving you a single model to use, we just had you compare and evaluate two differing models of charge. What is the value, if any, in taking this approach?*

---

While the content of this ILD focuses on physics rather than economics concepts, the systematic process of starting with students' mental models developed from previous experience (step 1), prompting for alternative models (step 2), using activities to generate data that can be used to evaluate competing models (step 3), and asking students to think intentionally about their thinking process in carrying out the ILD (steps 4 and 5), is universally applicable. The activities that the students are engaged in are not simply aimed at getting a correct answer or learning a specific concept (although these are important) but are meant to put into practice the key learning attributes discussed earlier in this paper.

In economics, classroom experiments are perhaps the most similar to ILDs. However, few classroom experiments in economics are as intentional as ILDs in tying the experiment to students' initial experiences, specifically targeting student misconceptions, and helping students to think about their learning in a metacognitive manner. Instructors using economic experiments could benefit by exploring the framework developed for physics experiments and adapting it for use with economic experiments in the classroom.

In the following example we illustrate how the ILD approach can be adapted for use in the economics principles course to address a common student confusion – the difference between price levels and inflation. As recommended by learning theory, the starting point for the activity is students' current understanding – including misconceptions – of the concept. Students are first asked to describe their experiences with inflation. In our experience students often focus their attention on "high" prices for individual items rather than changes in the overall price level in the economy. The activity is aimed at creating dissonance between their initial naïve conception of the topic and subsequent understanding, confronting the reality that inflation is not the same as high prices.

In addition, students are prompted to be metacognitive, providing practice in monitoring their own thinking processes, and in the final step, are asked to explain why there often is confusion on the core concept being examined. In addition, students are asked to transfer their understanding between the steepness of a hill, something that they are all familiar with, and the rate of change in a price index. Finally, they are introduced to the idea of using economic data to develop empirical descriptions of economic phenomena.

---

**ILD Example – Economics**

*Defining Concepts Carefully:
Thinking about Inflation*

*1. Personal Experiences*

*What experiences have you had that you attribute to inflation? Considering these experiences, can you generate a description of inflation that we can use in discussing economic models of the economy this semester?*

*2. Modeling Skills Practice*

*Imagine that you are climbing the hill shown in the foreground of the picture below.*

- *When is your climb the steepest?*
- *Is it necessarily when you are at the top of the hill? Why not?*
- *What is the difference between steepness and height?*

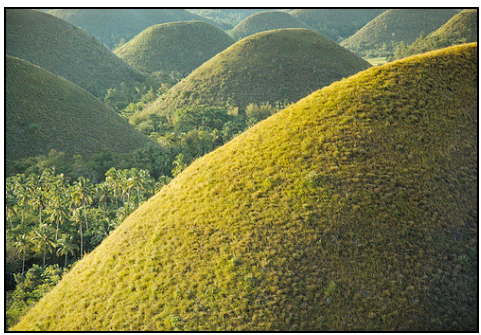

Source: http://flickr.com/photos/jan2you/515005127/

### 3. Experiment/Demonstration

*Now let's transfer this concept to a graph of economic data. The graph below illustrates the consumer price index (CPI), the basic measure of the price level in the economy, over the period 1970 – 1986.*

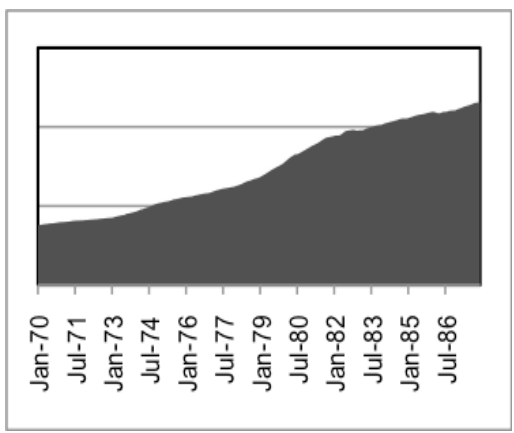

Source: http://alfred.stlouisfed.org/series?seid=CPIAUCNS

*Given the data above on the Consumer Price Index:*

- *When is the change in the average price level in the economy the steepest?*
- *Is it necessarily when you are at the top of the graph? Why not?*
- *What is the difference between steepness and height in this graph?*

### 4. Resolution

*In what ways do your responses to the questions in questions 2. and 3. above support or conflict with your description of your experience with inflation in question 1.?*

### 5. Reflection (Metacognition)

*Why do people confuse high prices with inflation?*

The example above parallels the structure of the physics ILD on electrical force but focuses on an important economics concept. However, note that while this ILD addresses an important student misconception, it risks leaving students with the idea that steepness in a graph is all that matters for measuring rates of change. Economic "experts" know that this is note true, recognizing that when dealing with percentage changes it is not the absolute change in a variable that is important, but the change *relative to the starting base*.

To move students to this more expert-like understanding, it may be necessary to first address the underlying misconception between price levels and rates of change. To extend the classroom ILD and address this issue, a follow-up exercise could be constructed to challenge the yet-naïve belief that only steepness matters. Such a follow-up exercise could be conducted in class or assigned as an out-of-class activity. In this way, ILDs can help "scaffold" student learning, beginning with simple (yet important for student learning) ideas and concepts and intentionally building students' thinking skills through a series of intentionally-designed activities.

Other economics ILD examples are possible, such as getting students to consider the source of changes in the price of a specific good, focusing on the relative roles of demand and supply shifts in causing price changes. It would be useful to have a library of economics ILD exercises aimed at addressing student misconceptions for the most troublesome concepts in economics and to test the efficacy of these exercises across institutions, instructors, and students. Such a library could be developed and added to through a focused effort of a team of economic education researchers.

# IV. SUMMARY

As noted in the introduction, we believe that economists can learn a great deal from educational research practices and related pedagogical innovations in other disciplines, in particular physics education. Physics shares many of the educational challenges faced by economists and has developed a deep research-grounded educational knowledge base that serves as a foundation for a growing number of effective pedagogical practices. These physics education pedagogical innovations, in turn, provide fertile opportunities for adaptation and testing in economics, while the physics education research framework serves as an alternative lens through which to view student learning.

The first part of this paper outlined some of the defining characteristics of physics education research as a means of highlighting key differences between the way that economists and physicists undertake educational research in their fields. Three characteristics are particularly important in this respect: (1) the intentional grounding of physics education research in learning science principles, (2) a shared conceptual research framework focused on *how* students learn physics concepts, and (3) a cumulative process of knowledge-building in the discipline. These three aspects have enabled physics education research to progress rapidly during the past thirty years, leading to a series of pedagogical and curricular innovations that are well known within physics community. Similar conceptual and procedural coherence is generally lacking in economics education.

We believe that understanding the differences in research style between physics and economics education researchers is important for developing effective pedagogical innovations that begin to directly address student learning difficulties in economics. Physics education researchers have spent the better part of thirty years researching exactly this question. To make these ideas more concrete for economists, in the second half of the paper we identified four specific physics-education-research-inspired teaching innovations that we believe hold particular promise for improving student learning in economics, and illustrated how they might be implemented in economics courses. Our examples are intended to be illustrative, rather than comprehensive, offering a small "starter sample" to highlight how pedagogical innovations developed in one discipline can be successfully adapted for use in another.

What distinguishes these four examples from similar work in economics is the intentionality of their development and use, in particular their grounding in learning science principles and their linkage to the cumulative body of physics education research. Economists may *do* similar things in their own classrooms, but for different reasons. In physics, the intentional development and implementation of pedagogical innovations has been shown to be effective at increasing students' physics knowledge. Would we get similar results in economics by following similar strategies? What is needed now is additional empirical research to answer this question and determine how these tools can best be used in economics to uncover and address student misconceptions, promote reflective learning, and develop students' ability to transfer knowledge to new situations. By looking outside our own discipline for answers to these questions, we are likely to gain valuable insights about how students learn economic concepts, adding to a cross-disciplinary cumulative knowledge base that is likely to spawn additional pedagogical innovations aimed at improving student learning.